\documentclass[a4paper,11pt]{article}

\usepackage{jheppub} 
\usepackage[T1]{fontenc} 
\usepackage[italian,english]{babel}
\usepackage{hyperref}
\usepackage{ifpdf}
\usepackage{subfigure}
\usepackage{tikz}
\usepackage[compat=1.1.0]{tikz-feynman}
\usepackage{slashed}
\usetikzlibrary{arrows,shapes}
\usetikzlibrary{trees}
\usetikzlibrary{matrix,arrows} 				
\usetikzlibrary{positioning}				
\usetikzlibrary{calc,through}				
\usetikzlibrary{decorations.pathreplacing}  
\usepackage{pgffor}							

\usetikzlibrary{decorations.pathmorphing}	
\usetikzlibrary{decorations.markings}
\tikzset{
	vector/.style={decorate, decoration={snake}, draw},
	fermion/.style={draw=black, postaction={decorate}}, 
	scalar/.style={dashed,draw=black, postaction={decorate}}}
\tikzstyle{block} = [draw, rectangle, 
minimum height=3em, minimum width=6em]

\usepackage{amssymb}
\usepackage{amsfonts}
\usepackage{epsf}
\usepackage{rotating}
\usepackage{graphicx}
\usepackage{amsmath}
\usepackage{fancyhdr}
\usepackage{lineno}
\usepackage{babel}
\usepackage{graphics}
\usepackage{pstricks}
\usepackage{color}
\usepackage{multirow}
\usepackage{float}
\usepackage{lineno}
\usepackage{mathtools}
\usepackage{stackengine}
\newcommand{\be}{\begin{equation}}
\newcommand{\ee}{\end{equation}}
\newcommand{\bea}{\begin{eqnarray}}
\newcommand{\eea}{\end{eqnarray}}

\catcode`@=12

\preprint{CQUeST-2025-0757}

\title{\boldmath Scrutinizing Fermionic Dark Matter in Scotogenic Model with Low Reheating Temperature}

\author[a,b]{Abhishek Roy,}
\author[c,d]{Rameswar Sahu}

\affiliation[a]{Center for Quantum Spacetime, Sogang University, 35 Baekbeom-ro, Mapo-gu, Seoul, 121-742, South Korea}
\affiliation[b]{Department of Physics, Sogang University, 35 Baekbeom-ro, Mapo-gu, Seoul, 121-742, South Korea}
\newcommand{\AddrHBNI}{
	Homi Bhabha National Institute, BARC Training School Complex, Anushakti Nagar, Mumbai 400094, India }
\affiliation[c]{Institute of Physics, Sachivalaya Marg, Bhubaneswar, Odisha 751005, India}
\affiliation[d]{\AddrHBNI}

\emailAdd{abhishek@sogang.ac.kr}
\emailAdd{rameswar.s@iopb.res.in}

\abstract{
The scotogenic model provides a minimal and elegant framework that simultaneously explains neutrino masses and accommodates a viable dark matter (DM) candidate. In this work, we investigate the phenomenology of fermionic DM in the scotogenic model, with a particular emphasis on the effects of a non-standard cosmological history characterized by a low reheating temperature. We demonstrate that entropy injection from inflaton decay can significantly dilute the DM abundance, thereby relaxing the annihilation cross section required to reproduce the observed relic density and opening new regions of viable parameter space. We further analyze the complementarity between current and future direct detection experiments and charged lepton flavour violation (cLFV) searches in probing this scenario. Our results show that next-generation direct detection experiments such as DARWIN and XLZD, together with upcoming cLFV searches (in particular the future sensitivity of $\mu \rightarrow 3e$ and $\mu \rightarrow e$ conversion experiments), will be capable of testing substantial regions of the parameter space, including those associated with low reheating temperatures. 
}
\begin{document} 

\maketitle
\flushbottom

\section{Introduction}
   
The Standard Model (SM) of particle physics, while standing as a remarkable achievement of human intellect and experimental ingenuity, remains incomplete. Phenomena such as the observed matter--antimatter asymmetry \cite{WMAP:2010qai}, the discovery of neutrino flavour oscillations implying tiny but non-zero neutrino masses \cite{SNO:2002tuh, SNO:2002hgz, KamLAND:2002uet, K2K:2002icj, Super-Kamiokande:2006jvq, T2K:2011ypd}, and the compelling evidence for dark matter (DM) from astrophysical and cosmological observations \cite{Rubin:1970zza, Fixsen:1996nj, Planck:2018vyg, Primack:1997av, Clowe:2003tk}, all lie beyond the explanatory reach of the SM. These observations strongly suggest the existence of physics beyond the Standard Model (BSM). Among the many proposed extensions, those capable of addressing multiple fundamental puzzles in a unified framework are especially compelling \cite{Krauss:2002px,Asaka:2005an,Ma:2006km,Aoki:2008av,Restrepo:2013aga,Escudero:2016ksa,Cai:2017jrq,Cacciapaglia:2020psm}, and have been the subject of intense theoretical and experimental scrutiny. In this work, we focus on one such scenario: the scotogenic model \cite{Tao:1996vb,Ma:2006km,Avila:2025qsc}, which elegantly accounts for the origin of neutrino masses while simultaneously providing a cosmologically viable dark matter candidate.

The scotogenic model extends the Standard Model by introducing an inert scalar doublet $\eta$ and three singlet fermions $N_i$ ($i=1,2,3$), all of which are odd under a discrete $Z_2$ symmetry. This symmetry not only forbids the generation of neutrino masses at tree level but also ensures the stability of the lightest $Z_2$-odd particle, thereby providing a natural dark matter candidate. Neutrino masses are generated radiatively at the one-loop level, with dark sector particles propagating in the loop. Depending on the spectrum, the dark matter can be either scalar or fermionic in nature.  

The scalar dark matter scenario is comparatively more accessible, primarily because it can interact through Higgs-portal–like couplings. Consequently, it has been extensively studied in the literature, with numerous analyses exploring the parameter space consistent with the observed relic density and subjecting it to scrutiny through direct detection \cite{Dolle:2009fn,Arhrib:2013ela,Belyaev:2016lok}, indirect detection \cite{Queiroz:2015utg,Garcia-Cely:2015khw,Eiteneuer:2017hoh}, and collider searches \cite{Dolle:2009ft,Miao:2010rg,Kalinowski:2018kdn,Yang:2021hcu,Fan:2022dck}. In contrast, the fermionic dark matter scenario poses greater challenges. The scattering of the fermionic candidate with nucleons is loop-induced \cite{Schmidt:2012yg,Ibarra:2016dlb}, leading to a suppressed direct detection signal. Furthermore, owing to the Majorana nature of the lightest fermion $N_1$ \cite{Kubo:2006yx}, its annihilation cross section is $p$-wave suppressed, rendering indirect detection prospects less promising. Collider constraints on the scotogenic model mainly arise from LHC searches for di-lepton + missing transverse momentum final states, originating from pair production of the charged inert scalars followed by their decays into SM leptons and singlet fermions $N_i$. A recent reinterpretation study \cite{Baumholzer:2019twf} of ATLAS Run--2 searches \cite{ATLAS:2018ojr, ATLAS:2017qwn} at $\sqrt{s}=13~\mathrm{TeV}$ with $36~\mathrm{fb}^{-1}$ of data shows that the current LHC dataset does not yet place robust exclusions on this scenario. 

Projections for the High-Luminosity LHC at $\sqrt{s}=14~\mathrm{TeV}$ with $4~\mathrm{ab}^{-1}$ indicate sensitivity only in a limited region, typically up to charged scalar masses of ${\cal O}(400$--$600)\,\mathrm{GeV}$ and singlet fermion masses of a few hundred~GeV, depending on the Yukawa structure and mass splittings \cite{Baumholzer:2019twf}. In contrast, future facilities such as the future circular hadron collider (FCC-hh) \cite{FCC:2018vvp} at $\sqrt{s}=100~\mathrm{TeV}$ with $3$--$30~\mathrm{ab}^{-1}$ and the Compact Linear Collider (CLIC) \cite{CLIC:2018fvx, CLIC:2016zwp} at $\sqrt{s}=3~\mathrm{TeV}$ with $3~\mathrm{ab}^{-1}$ are expected to significantly extend the coverage, probing TeV-scale inert scalars and singlet fermions (up to $\sim 1$--$2~\mathrm{TeV}$ in favorable regions of parameter space) \cite{Baumholzer:2019twf}. In this paper our focus will be on the fermionic DM scenario of the scotogenic model.

One of the central aims of this work is to explore the prospects of probing the fermionic dark matter scenario through both current and future direct detection experiments, as well as via charged lepton flavour violating (cLFV) observables. Although the dark matter--nucleon interaction is loop suppressed, there exist regions of parameter space where the scattering cross section can be significantly enhanced. For instance, near-degeneracy among the singlet fermions \cite{Schmidt:2012yg} or the presence of a sizable quartic coupling between the inert scalar doublet and the SM Higgs doublet \cite{Ibarra:2016dlb} can yield a relatively large spin-independent cross section, thereby rendering the scenario testable in upcoming direct detection searches. In parallel, cLFV processes such as $\mu \to e \gamma$, $\mu \to 3e$, or $\mu \rightarrow e$ conversion arise naturally in this framework and provide stringent constraints on the Yukawa structure of the model \cite{Toma:2013zsa, Vicente:2014wga}. Anticipated improvements in the sensitivity of cLFV experiments therefore hold the potential to probe regions of parameter space that may remain inaccessible to direct detection, highlighting the complementarity of these two avenues in testing the fermionic dark matter scenario.

Another key objective of this work is to investigate the impact of a low reheating temperature on dark matter phenomenology within the scotogenic model. Most previous studies have assumed the standard cosmological history, where reheating after inflation is effectively instantaneous. In that case, the reheating temperature is always higher than the temperature at which dark matter freezes out, and thus the relic density calculation follows the conventional thermal scenario. Recent developments, however, have highlighted that if the reheating temperature is lower than the dark matter decoupling temperature, the late decay of the inflaton can inject entropy into the thermal bath during the reheating era \cite{Allahverdi:2020bys, Batell:2024dsi,Silva-Malpartida:2023yks,Haque:2023yra,Cosme:2023xpa,Cosme:2024ndc,Bhattiprolu:2022sdd}. This modifies the standard relic abundance calculation, as the additional entropy dilutes the dark matter density and consequently lowers the annihilation cross section required to reproduce the observed relic abundance \cite{Belanger:2024yoj, Gelmini:2006pw, Bernal:2022wck, Bernal:2024yhu,Mondal:2025awq}. The reheating temperature itself is determined by the inflaton decay width and can, in principle, take a wide range of values \cite{Sarkar:1995dd, Kawasaki:2000en, Hannestad:2004px, DeBernardis:2008zz, deSalas:2015glj}, provided it remains above the lower bound set by Big Bang nucleosynthesis (BBN). Such a non-standard cosmological history opens up new regions of parameter space that would otherwise be excluded in the standard picture, thereby offering fresh perspectives on the viability of fermionic dark matter in the scotogenic framework.

The remainder of this paper is organized as follows. In Section~\ref{sec:model}, we briefly review the scotogenic model and its key features. Section~\ref{sec:DMobs} is devoted to the phenomenology of fermionic dark matter in this framework, with particular emphasis on the implications of a low reheating temperature. In Section~\ref{sec:constraints}, we outline the theoretical and experimental constraints that shape the viable parameter space of the model. Our main results are presented in Section~\ref{sec:results}. Finally, in Section~\ref{sec:conclusion}, we summarize our findings and discuss possible directions for future investigation.

\begin{table}[]
	\centering
	\begin{tabular}{|c|c|c|c|c|c|c|}
		\hline \phantom{XXXXXXXX} & \phantom{X} $L_i$ \phantom{X} & \phantom{X}$\ell_{R_i}$\phantom{X} & \phantom{X} $\Phi$ \phantom{X} & \phantom{X} $\eta$ \phantom{X} & \phantom{X} $N_i$ \phantom{X}  \\
		\hline \hline
		{\bf  $SU(2)_L$} & {\bf  $2$} & {\bf  $1$} & {\bf  $2$} & {\bf  $2$} & {\bf  $1$}  \\
		\hline
		{\bf  $U(1)_Y$} & {\bf  $-1/2$} & {\bf  $-1$} & {\bf  $1/2$} & {\bf  $1/2$} & {\bf  $0$} \\
		\hline
		$\mathbb{Z}_2$ & $+$ & $+$ & $+$  & $-$ & $-$ \\
		\hline
	\end{tabular}
	\caption{Particle content and their charge assignments under various symmetry groups.}
	\label{tab:particle-content}
\end{table}

\section{The Model}
\label{sec:model}
The Scotogenic model extends the SM by introducing $SU(2)_{L}$ scalar doublet $\eta$, with hypercharge $Y = 1/2$ and three SM gauge singlet fermions, $N_{i}$, where the index $i$ runs from 1 to 3. We show the particle content and their quantum numbers under the SM gauge group  $SU(3)_c\otimes SU(2)_L\otimes U(1)_Y$ in Table~\ref{tab:particle-content}. The Lagrangian features a discrete $\mathbb{Z}_2$ symmetry alongside the gauge symmetries of the SM. The $\mathbb{Z}_2$ symmetries prevent flavour-changing neutral currents but also ensure the lightest odd particle in the spectrum is stable, making it a viable dark matter (DM) candidate. Consequently, the model accommodates two plausible DM scenarios—one involving a scalar and the other a fermion. In this work, we focus on the case where DM is fermionic. The renormalizable gauge-invariant terms in the Lagrangian responsible for neutrino mass generation are given by,

\begin{equation}
-\mathcal{L}_{\text{Y}} \supset Y^{\ell}_{\alpha i} \bar{L}_{\alpha} \Phi \ell_{R_i} + Y^{\nu}_{\alpha i} \, \bar{L}_{\alpha}  \tilde{\eta} N_i + \frac{1}{2} M_{N_k} \overline{N_k^c}N_k +\textit{h.c.},
\label{eq:LagYL}
\end{equation}
where $\tilde{\eta}=i\sigma_2\eta^*$, $L=(\nu_{\ell L},\ell_L)^T$, and we have assumed the mass matrix of the singlet fermions to be diagonal. The gauge invariant scalar potential  can be written as,
\begin{align}
V =  &m_\Phi^2 \Phi^\dagger \Phi + m_\eta^2 \eta^\dagger \eta + \frac{\lambda_1}{2} (\Phi^\dagger \Phi)^2 + \frac{\lambda_2}{2} (\eta^\dagger \eta)^2 + \lambda_3 (\eta^\dagger \eta)(\Phi^\dagger \Phi) + \lambda_4 (\eta^\dagger \Phi) (\Phi^\dagger \eta )+ \nonumber \\  &\frac{\lambda_5}{2} \left[ (\eta^\dagger \Phi)^2 + \text{h.c} \right].\label{eq:potential}
\end{align}
In the limit $\lambda_5 \to 0$, the global symmetry associated with lepton number is restored, and radiative neutrino masses vanish (see the following discussion). Thus, $\lambda_5$ controls the amount of lepton–number violation relevant for neutrino mass
generation in this framework. For our analysis, we have assumed all the parameters of the scalar potential to be real. Additionally, we require $m_\eta^2 > 0$ to prevent breaking the $Z_2$ symmetry. We expand the fields $\Phi$ and $\eta$ as follows,
\begin{align}
\Phi=
\begin{pmatrix}
\phi^+\\
(v_\Phi+h+i\eta_\Phi)/\sqrt{2}
\end{pmatrix} \,\,\text{and}\,\,
\eta=
\begin{pmatrix}
\eta^+\\
(\eta^R+i\eta^I)/\sqrt{2}
\end{pmatrix}.
\label{eqmass}
\end{align}
In the unitary gauge, the would-be Goldstone bosons $\eta_\Phi$ and $\phi^+$ from the SM Higgs doublet $\Phi$ are absorbed by the SM gauge bosons $Z$ and $W^\pm$, respectively, thus generating their masses. The mass of the SM-like Higgs boson, denoted by $h$, is given by
\begin{align}
m_{h}^2=\lambda_{1}v_{\Phi}^2,
\end{align}
where $v_{\Phi} = 246\, \text{GeV}$ denotes the vacuum expectation value (VEV) of the SM Higgs field. It's important to note that the mixing between the Higgs field $\Phi$ and the dark doublet $\eta$ is forbidden due to the exact preservation of the $\mathbb{Z}_2$ symmetry. The components of $\eta$ have the following masses:

\begin{align}
m_{\eta^{+}}^{2}&=m_{\eta}^{2}+\frac{1}{2}\lambda_{3}v_\Phi^{2},\\
m_{\eta^{R}}^{2}&=m_{\eta}^{2}+\frac{1}{2}\lambda_{L}v_\Phi^{2},\\
m_{\eta^{I}}^{2}&=m_{\eta}^{2}+\frac{1}{2}\bar{\lambda}_{L}v_\Phi^{2},
\end{align}
where we have defined
\begin{align}
\lambda_{L}=\left(\lambda_{3}+\lambda_{4}+\lambda_{5}\right)\,\,\,\text{and} \,\,\, \bar{\lambda}_{L}= \left(\lambda_{3}+\lambda_{4}-\lambda_{5}\right).
\end{align}

Starting from the scalar potential defined in Eq.~\ref{eq:potential}, we observe that the potential contains seven free parameters. Since the field $\Phi$ plays the same role as the SM Higgs doublet, the parameters $m_\Phi$ and $\lambda_1$ are fixed by the measured Higgs boson mass and electroweak precision data. Among the remaining parameters, $\lambda_2$ does not influence the scalar mass spectrum at tree level. However, as we will elaborate in the subsequent sections, $\lambda_2$ plays a pivotal role in ensuring the stability of the electroweak vacuum. For the sake of simplicity, we fix $\lambda_2 = 0.1$ throughout our analysis.

The parameters $\lambda_3$ and $\lambda_4$ play an important role in dark matter phenomenology, particularly with respect to direct detection (DD) constraints. In our analysis, we vary them within suitable ranges to ensure consistency with both theoretical requirements and electroweak precision observables, namely the oblique parameters $S$ and $T$. The remaining free parameters in the scalar potential are $m_{\eta}$ and $\lambda_5$. However, in our study, instead of treating $m_{\eta}$ as a free parameter, we take $m_{\eta_R}$ as independent.

\par
\begin{figure}[]
	\centering
	\includegraphics[height=5cm,width=9cm]{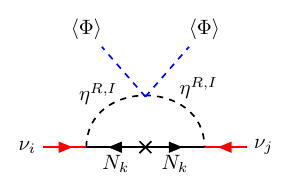}
	\caption{One loop Feynman diagram for neutrino mass generation.}
	\label{fig:loop-neutrino}
\end{figure}

 Although the usual tree-level seesaw contribution to neutrino masses is forbidden by the $\mathbb{Z}_2$ symmetry, these are induced at the 1-loop through the exchange of the ``dark'' fermions and scalar as illustrated in Fig.~\ref{fig:loop-neutrino}. This loop is calculable and neutrino mass is given by the following expression \cite{Mandal:2021yph,Ma:2006km,DeRomeri:2022cem,Vicente:2014wga,Chun:2023vbh}~\footnote{Note that the neutrino mass expressions presented in Refs.~\cite{Vicente:2014wga,Ma:2006km} differ from Eqn.~\ref{eq:numass} by an overall factor of 1/2. The expression in Eqn.~\ref{eq:numass} follows the corrected convention, as adopted in Refs.~\cite{DeRomeri:2022cem,Mandal:2021yph,Chun:2023vbh}.}
\begin{align}
\label{eq:numass}
(m_\nu)_{ij} &
= \sum_{k=1}^3 \frac{Y^\nu_{ik} \, Y^\nu_{kj} M_{N_k}}{32 \pi^2}\left[ \frac{m_{\eta_R}^2}{m_{\eta_R}^2-M_{N_k}^2} \log \frac{m_{\eta_R}^2}{M_{N_k}^2} -  \frac{m_{\eta_I}^2}{m_{\eta_I}^2 - M_{N_k}^2} \log \frac{m_{\eta_I}^2}{M_{N_k}^2} \right],\\
& \equiv (Y^\nu{^T} \Lambda Y^{\nu})_{ij},
\end{align}
where $\Lambda$ matrix is defined as $\Lambda=\text{diag}(\Lambda_1,\Lambda_2,\Lambda_3)$, with
\begin{align}
\label{eq:Lambda}
\Lambda_k
=  \frac{M_{N_k}}{32 \pi^2}\left[ \frac{m_{\eta_R}^2}{m_{\eta_R}^2-M_{N_k}^2} \log \frac{m_{\eta_R}^2}{M_{N_k}^2} -  \frac{m_{\eta_I}^2}{m_{\eta_I}^2 - M_{N_k}^2} \log \frac{m_{\eta_I}^2}{M_{N_k}^2} \right].
\end{align}

It is important to highlight that in the limit $\lambda_5 \to 0$, the scalar masses satisfy $m_{\eta^R}^2 = m_{\eta^I}^2$. This mass degeneracy leads to an exact cancellation between the loop contributions from $\eta^R$ and $\eta^I$, resulting in vanishing neutrino masses. Thus, the parameter $\lambda_5$ plays a critical role in generating small but non-zero neutrino masses in the Scotogenic model.

To systematically explore the implications of neutrino data, we adopt the Casas-Ibarra parametrization~\cite{Casas:2001sr} for the Yukawa coupling matrix:
\begin{align}
Y^\nu = \sqrt{\Lambda}^{-1} R \sqrt{\hat{m}_\nu} U^{\dagger}_{\rm lep},
\label{eq:Ynu}
\end{align}
Here, $R$ is a general $3 \times 3$ complex orthogonal matrix, $U_{\rm lep}$ denotes the Pontecorvo-Maki-Nakagawa-Sakata (PMNS) matrix responsible for diagonalizing the neutrino mass matrix, and $\hat{m}_\nu$ is the diagonal matrix of light neutrino masses. The matrix $R$ can be expressed as:
\begin{align}
R = \left( \begin{array}{ccc}  1 & 0 & 0
\\ 0 & \cos x & \sin x \\
0 & -\sin x & \cos x\\
\end{array} \right)  \left( \begin{array}{ccc}  \cos y & 0 & \sin y
\\ 0 & 1 & 0 \\
-\sin y & 0 & \cos y\\
\end{array} \right)  \left( \begin{array}{ccc}  \cos z & \sin z & 0
\\ -\sin z & \cos z & 0 \\
0 & 0 & 1\\
\end{array} \right),
\label{eq:R_matrix}
\end{align}

For the purposes of the present work, we adopt a minimal and controlled setup by fixing the matrix $R$ to be the Identity matrix, corresponding to $x = y = z = 0$. This simplifying and conservative choice allows us to assess whether the model can simultaneously satisfy the dark matter, neutrino, and current cLFV constraints already in a highly restricted parameter space, without invoking additional freedom from complex rotations in the Yukawa sector. We emphasise that this assumption is not intended to be general: allowing the complex angles $x$, $y$, $z$ to vary can enlarge the viable parameter space and may have a significant impact on cLFV observables, as discussed for example in Ref.~\cite{Karan:2023adm}. A comprehensive exploration of the fully unrestricted Casas--Ibarra parameter space, with both real and imaginary components of the angles varied, lies beyond the scope of the present work and will be addressed in a forthcoming dedicated study. Furthermore, we work with the normal ordering (NO) of neutrino masses and fix the neutrino oscillation parameters to their current best-fit values as reported in Ref.~\cite{deSalas:2020pgw}, as summarized in Table~\ref{tab:Neutrino_Best_Fit}.

\begin{table}[]
	\centering
\begin{tabular}{|c|c|}
	\hline
	parameter & best fit  \\
	\hline
	$\Delta m^2_{21} [10^{-5}\,\text{eV}^2]$  
	& $7.5$  \\[4mm]
	$|\Delta m^2_{31}| [10^{-3}\,\text{eV}^2]$ (NO)  
	& $2.55$  \\[4mm]
	$\sin^2\theta_{12} / 10^{-1}$  
	& $3.18$  \\
	$\theta_{12} /^\circ$  
	& $34.3$  \\[4mm]
	$\sin^2\theta_{23} / 10^{-1}$ (NO)  
	& $5.74$  \\
	$\theta_{23} /^\circ$ (NO)  
	& $49.26$  \\[4mm]
	$\sin^2\theta_{13} / 10^{-2}$ (NO)  
	& $2.2$  \\
	$\theta_{13} /^\circ$ (NO)  
	& $8.53$  \\[4mm]
	$\delta/\pi$ (NO)  
	& $1.08$  \\
	$\delta/ ^\circ$ (NO)  
	& $194$  \\[3mm]
	\hline
\end{tabular}
	\caption{Best-fit values of neutrino oscillation parameters from a global analysis Ref.~\cite{deSalas:2020pgw}.}
	\label{tab:Neutrino_Best_Fit}
\end{table}

\section{Dark Matter Observables}
\label{sec:DMobs}
In this work, we consider the fermionic dark matter realization of the scotogenic model, where the lightest $Z_2$-odd fermion $N_1$ serves as the WIMP dark matter candidate. Our primary objective is to investigate how a low reheating temperature modifies the dark matter phenomenology in this framework.

To isolate the impact of the reheating dynamics, we choose a region of parameter space in which \emph{scalar--DM co-annihilation} processes are strongly suppressed. This is implemented by enforcing a mass hierarchy of the form
\begin{equation}
    m_{\eta^+,\eta^R,\eta^I} \;>\; 1.5\, M_{N_1},
\end{equation}
so that the inert scalar states are sufficiently heavier than $N_1$ and their thermal number densities are Boltzmann suppressed at freeze-out. Consequently, co-annihilation channels involving the scalar fields do not play a significant role in determining the relic abundance~\cite{Molinaro:2014lfa}. In this regime, the dominant annihilation processes relevant for the freeze-out dynamics are those mediated by the inert scalars and involving the fermionic states of the model.

It is important to emphasize that this assumption does \emph{not} exclude the possibility of \emph{fermion--fermion co-annihilation} among $N_1$, $N_2$, and $N_3$. In particular, for parameter points where the masses of these fermions are quasi-degenerate, such channels can contribute to the effective annihilation rate. All kinematically allowed annihilation and co-annihilation processes among $N_1$, $N_2$, and $N_3$ are consistently incorporated in our computation of the thermally averaged cross section using \texttt{micrOMEGAs} \cite{Alguero:2023zol,Belanger:2024yoj}. The resulting temperature-dependent quantity $\langle \sigma v \rangle (T)$, which by construction includes the fermionic co-annihilation contributions, is then employed in our numerical solution of the Boltzmann equation in the low-reheating-temperature scenario studied here.

\subsection{Relic Density}

The relic abundance of dark matter in the early universe is governed by its interaction rate with the thermal plasma. For the fermionic WIMP candidate $N_1$ in the scotogenic model, this evolution is described by the Boltzmann equation~\cite{Kolb:1988aj},
\begin{equation}
    \frac{dn}{dt} + 3Hn
    = -\langle \sigma v \rangle
    \left(n^2 - n_{\text{eq}}^2\right),
    \label{Eqn:Beq1_N1}
\end{equation}
where $n$ is the total number density of the co-annihilating fermionic species, $H$ is the Hubble expansion rate, and $\langle \sigma v \rangle$ denotes the thermally averaged annihilation cross section that includes both annihilation and fermion--fermion co-annihilation channels when the mass splittings are small.

In the region of parameter space where $N_1$ is the only thermally populated $Z_2$-odd species, the dominant contribution to $\langle \sigma v \rangle$ arises from $N_1 N_1 \to \ell^{\pm}\ell^{\mp},\,\nu_\ell\nu_\ell$, mediated by the inert scalars $\eta^{\pm}$ and $\eta^{R,I}$. In this limit, the thermal averaged annihilation cross section can be written as $\langle \sigma v \rangle = A + B v^2$, with~\cite{Ibarra:2016dlb,Kubo:2006yx},
\begin{equation}
    A = 0, \quad
    B = \frac{M_{N_1}^2 y_1^4}{48\pi}
    \left[
        \frac{M_{N_1}^4 + m_{\eta^+}^4}{(M_{N_1}^2 + m_{\eta^+}^2)^2}
        +
        \frac{M_{N_1}^4 + m_{\eta_{R,I}}^4}{(M_{N_1}^2 + m_{\eta_{R,I}}^2)^2}
    \right],
\end{equation}
where $y_1^2 = \sum_{\alpha} |Y^\nu_{1\alpha}|^2$ encodes the relevant Yukawa couplings. The corresponding equilibrium abundance is given by~\cite{Kolb:1988aj},
\begin{equation}
    n_{\text{eq}}(T)
    = \frac{g}{2\pi^2}\,M_{N_1}^2 T\,
      K_2\!\left(\frac{M_{N_1}}{T}\right),
\end{equation}
with $g = 2$ for a Majorana fermion and $K_2$ the modified Bessel function of the second kind.


To capture the effects of a low reheating temperature, we solve the Boltzmann equation in a cosmological background, dominated initially by a decaying inflaton field. To solve Eqn.~\ref{Eqn:Beq1_N1} in terms of the scale factor $a$, we define comoving yield $z=na^3$, which transforms it into:
\begin{equation}
    \frac{dz}{da} = -\frac{1}{H a^4} \langle \sigma v \rangle \left(z^2 - \tilde{z}^2\right),
\end{equation}
where $\tilde{z} = n_{\text{eq}} a^3$ is the comoving equilibrium yield. Furthermore, the DM relic density can be derived by,
\begin{equation}
    \Omega_{N_1}h^2=2.742\times10^8 m_{N_1}Y^0_{N_1},
\end{equation}
where $Y^0_{N_1}(=\frac{n_0}{s_{0}})$ is the asymptotic value of DM yield at the low temperature of the Universe, and $n_{0} $ and $s_0$ is the present-day number density and entropy density, respectively. The DM relic density has been precisely measured by the PLANCK satellite through observations of the cosmic microwave background (CMB)~\cite{Planck:2018vyg},
\begin{equation}
    \Omega^{obs}_{N_1}h^2=0.12\pm0.0012.
\end{equation}
In our numerical analysis, we will impose a $2\sigma$ limit on the DM relic density throughout our work.

The cosmological background is encapsulated by the the energy densities of the inflaton ($\rho_\phi$) and radiation ($\rho_R$) and evolve according to ~\cite{Silva-Malpartida:2023yks,Belanger:2024yoj},
\begin{align}
    \frac{d\rho_\phi}{da} + 3\frac{\rho_\phi}{a} &= -\frac{\Gamma_\phi}{H} \frac{\rho_\phi}{a}, \\
    \frac{d\rho_R}{da} + 4\frac{\rho_R}{a} &= \frac{\Gamma_\phi}{H} \frac{\rho_\phi}{a},
\end{align}
where $\Gamma_\phi$ denotes the inflaton decay width. The Hubble rate is sourced by both energy densities:
\begin{equation}
    H^2 = \frac{8\pi}{3M_P^2}(\rho_\phi + \rho_R),
\end{equation}
where $M_P=1.2\times10^{19}\,\text{GeV}$ is the Planck mass. For numerical integration, we switch to comoving variables $z_\phi = \rho_\phi a^3$ and $z_R = \rho_R a^4$, which satisfy:
\begin{align}
    \frac{dz_\phi}{da} &= -\frac{\Gamma_\phi}{H a} z_\phi,\label{eqn: backround_evolution1} \\
    \frac{dz_R}{da} &= \frac{\Gamma_\phi}{H} z_\phi.
    \label{eqn: backround_evolution2}
\end{align}
\begin{figure}[]
	\centering
	\includegraphics[height=6.5cm,width=7.2cm]{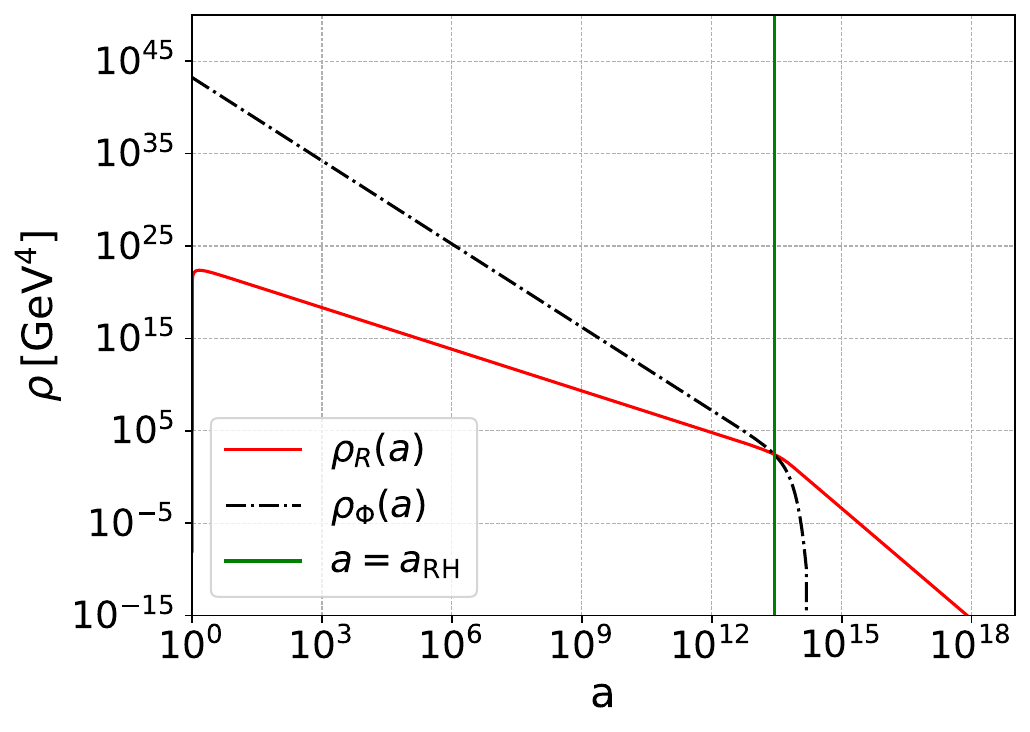}
	\includegraphics[height=6.5cm,width=7.2cm]{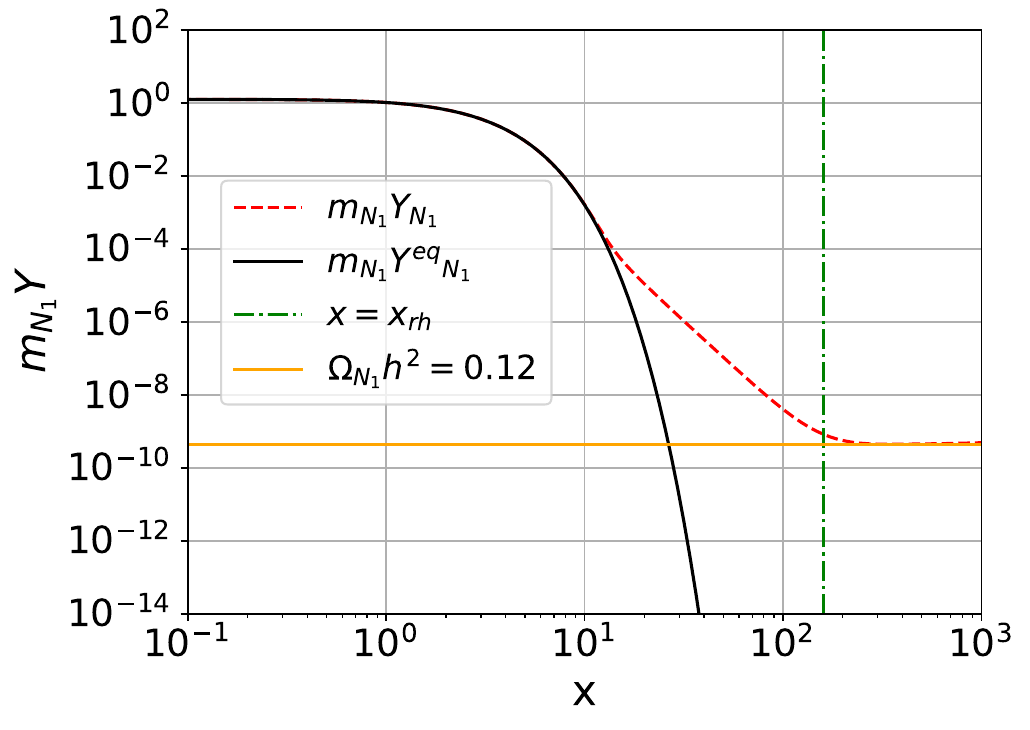}
	\caption{Left: Inflaton density ($\rho_{\phi}$) and radiation density ($\rho_R$) as functions of the scale factor $a$. Right: Dark matter yield ($m_{N_1}Y_{N_1}$) and equilibrium DM yield ($m_{N_1}Y^{\text{eq}}_{N_1}$) as functions of $x = m_{N_1}/T$. The vertical green line corresponds the reheating temperature, and the horizontal orange line denotes the observed relic density.}
	\label{fig:dm_evolution}
\end{figure}

The system is initialized at the onset of reheating ($a = 1$), where the radiation energy density is vanishing and the inflaton dominates:
\begin{equation}
    \rho_R(1) = 0, \quad \rho_\phi(1) = \frac{3M_P^2 H_I^2}{8\pi}.
\end{equation} 
where $H_I$ denotes the inflationary scale. Its maximum allowed value is constrained by the non-observation of primordial B-mode polarization in the cosmic microwave background, leading to an upper bound of approximately $4\times 10^{-6} M_P$ \cite{BICEP:2021xfz}. In our analysis, we fix the value of  $H_I$ at $10^3$ GeV, which is well within the observationally allowed range.  We begin by considering the evolution of the inflaton energy density ($\rho_{\phi}$) and the radiation energy density ($\rho_R$) as a function of the scale factor. The left panel of Fig.~\ref{fig:dm_evolution} displays our results for the benchmark scenario, i.e. $\Gamma_{\phi}=10^{-17}\,\rm{GeV}$ and $H_I=10^3\,\rm{GeV}$. Additionally, we indicate the scale factor corresponding to the reheating temperature, $a_{\text{RH}}$, as a vertical green line in the figure. The reheating temperature $T_{\text{RH}}$—defined by the condition $\rho_{\phi}(T_{\text{RH}}) = \rho_R(T_{\text{RH}})$—marks the onset of the radiation-dominated epoch. For the chosen benchmark scenario, $T_{\text{RH}}$ was numerically determined to be $1.8\,\mathrm{GeV}$. We further validated our numerical results by comparing them with those obtained using  micrOMEGAs~\cite{Alguero:2023zol,Belanger:2024yoj}, finding excellent agreement between the two.
\begin{figure}[]
	\centering
	\includegraphics[height=6.5cm,width=7.2cm]{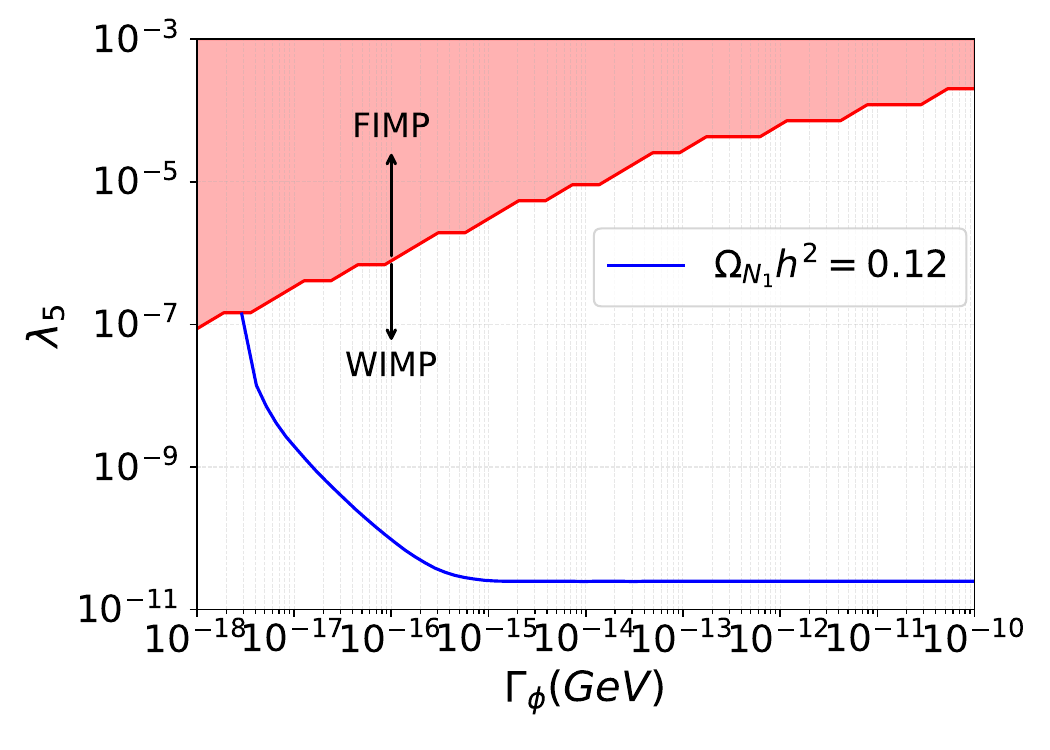}
    \includegraphics[height=6.5cm,width=7.2cm]{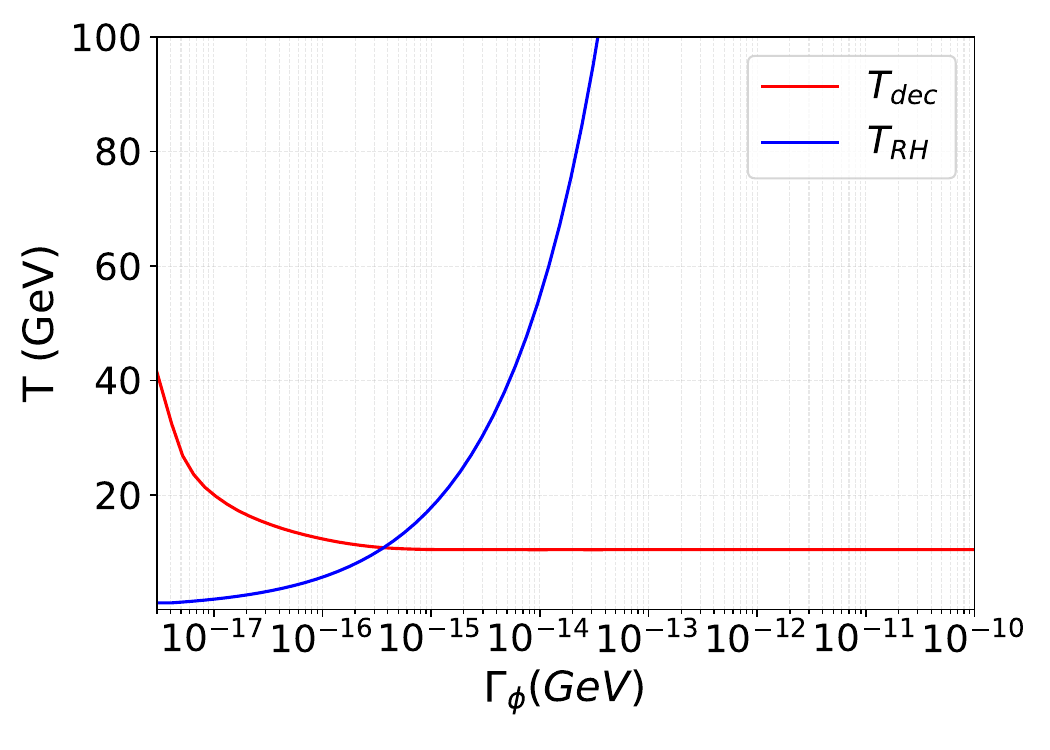}
	\caption{Left: Required values of $\lambda_5$ for different inflaton decay width $\Gamma_{\phi}$ in order to reproduce the observed dark matter relic density. The red line indicates the boundary separating the WIMP and FIMP regimes. Right: Evolution of the dark matter decoupling temperature and the reheating temperature as functions of $\Gamma_{\phi}$.}
	\label{fig:gammaphi_plots}
\end{figure}

In the right panel of Fig.~\ref{fig:dm_evolution}, we present the evolution of the DM yield, \( m_{N_1}Y_{N_1} \equiv m_{N_1}n/s \), defined as the ratio of the DM number density to the entropy density of the Universe, as a function of \( x = m_{N_1}/T \). The results corresponds to a fixed inflaton decay width \( \Gamma_{\phi} = 10^{-17}~\mathrm{GeV} \), taking into account the cosmological background evolution governed by Eqn.~\ref{eqn: backround_evolution1} and Eqn.~\ref{eqn: backround_evolution2}. For the model parameters, we have fixed $m_{\eta_R}$ at 763 GeV and $M_{N_1}$, $M_{N_2}$, $M_{N_3}$ at 289 GeV, 379 GeV, and 2615 GeV, respectively. The coupling \( \lambda_5 \) is chosen to be \( 3.68 \times 10^{-9} \), ensuring consistency with the observed DM relic abundance. Furthermore, the lightest neutrino mass is fixed at $7.23\times10^{-12}$ and the coupling $\lambda_3$ and $\lambda_4$ are fixed at $2.2\times 10^{-2}$ and $-2.5\times 10^{-3}$, respectively.

As evident from the figure, the DM co-moving number density experiences a dilution after it undergoes chemical decoupling, continuing until the Universe reaches the reheating temperature. This feature stands in stark contrast to the standard freeze-out scenario occurring during the radiation-dominated era, where the yield remains constant post-freeze-out. The observed dilution is a direct consequence of the entropy injection from the decaying inflaton field, characteristic of the low reheating temperature regime.

The left panel of Fig.~\ref{fig:gammaphi_plots} shows how the coupling \(\lambda_5\) must vary with the inflaton decay width \(\Gamma_{\phi}\) in order to reproduce the observed DM relic abundance. All other model parameters are fixed to the benchmark values used in Fig.~\ref{fig:dm_evolution}. The red shaded region corresponds to values of \(\lambda_5\) for which DM never reaches chemical equilibrium with the thermal bath, and is therefore FIMP-like. To aid interpretation of the dependence of \(\lambda_5\) on \(\Gamma_{\phi}\), the right panel of Fig.~\ref{fig:gammaphi_plots} displays the variation of the dark matter decoupling temperature ($T_{\rm{dec}}$) (defined by the condition, $(n_{eq}\langle\sigma v\rangle/H)_{T=T_{dec}}=1$) and the reheating temperature ($T_{RH}$) as functions of \(\Gamma_{\phi}\). As expected, the decay width \(\Gamma_{\phi}\) directly controls the duration of the reheating phase, with smaller values leading to lower reheating temperatures. When \(\Gamma_{\phi}\) falls below a certain threshold, the reheating temperature drops below the dark matter decoupling temperature, implying that the DM decouples from the SM thermal bath during the reheating-dominated era rather than in the standard radiation-dominated epoch.

As shown earlier in Fig.~\ref{fig:dm_evolution}, when decoupling takes place during the reheating era, the dark matter co-moving number density continues to be diluted after decoupling due to the sustained injection of entropy from inflaton decays. This post-decoupling entropy production effectively reduces the required thermally averaged annihilation cross-section to match the observed relic density. In the context of the Scotogenic model, where this cross-section is inversely proportional to \(\lambda_5\), a smaller cross-section implies that a larger value of \(\lambda_5\) is necessary. Consequently, \(\lambda_5\) must increase as \(\Gamma_{\phi}\) decreases in order to compensate for the dilution effect and reproduce the correct relic abundance. It is also important to note that for $\Gamma_\phi > 10^{-15}\,\mathrm{GeV}$—corresponding to $\lambda_5 \sim \mathcal{O}(10^{-10})$—the results reproduce the standard freeze-out scenario during the radiation-dominated era~\cite{Mandal:2021yph,Ma:2006km,Vicente:2014wga,Chun:2023vbh}.
\subsection{Direct Detection}
The spin-independent direct detection (SIDD) cross section of the fermionic DM candidate $N_1$ in the Scotogenic model has been computed in Refs.~\cite{Liu:2022byu, Ibarra:2016dlb}. In this work, we adopt the analytical expression provided in these references to constrain the viable parameter space of the model. The spin-independent scattering DM-nucleons cross section $\sigma_{SI}$ is given by~\cite{Liu:2022byu, Ibarra:2016dlb}:
\begin{equation}
    \sigma_{SI} = \frac{4}{\pi} \frac{M_{N_1}^2 m_p^4}{(M_{N_1}+m_p)^2} 
    \left( \frac{\Lambda_q}{m_q} \right)^2 f_p^2 \, ,
\end{equation}
where $m_p$ denotes the proton mass, $f_p\approx 0.3$ is the scalar form factor, and $\Lambda_q$ represents the effective scalar coupling, expressed as,
\begin{equation}
	\Lambda_q=-\frac{y_1^2}{16\pi^2 m_h^2 M_{N_1}}\left[\lambda_3 G_1\left(\frac{M_{N_1}^2}{m_{\eta^\pm}^2}\right)+\frac{\lambda_3+\lambda_4}{2}G_1\left(\frac{M_{N_1}^2}{m^2_{R,I}}\right)\right]m_q,
\end{equation}
with the loop function $G_1(x)$ given by
\begin{equation}
	G_1(x)=\frac{x+(1-x)\log(1-x) }{x}.
\end{equation}
Direct detection experiments place stringent upper bounds on $\sigma_{SI}$, thereby serving as powerful probes of WIMP scenarios such as the one considered here. In our analysis, we incorporate the experimental results from LUX-ZEPLIN (LZ)~\cite{LZ:2024zvo}, as it places more stringent limits compared to Xenon-1T~\cite{XENON:2018voc} and PandaX-4T ~\cite{PandaX-4T:2021bab}.
\begin{table}[]
	\centering
	\begin{tabular}{| c | c  | }
		\hline
		\textbf{Parameter} & \textbf{Scanned Range} \\
		\hline
		$m_{1}$ [GeV] & [$10^{-11}$ , $10^{-15}$] \\
        $M_{N_{1}}$ [GeV] & [$100$ , $500$] \\
        $m_{\eta_R}$ [GeV] & [$1.5M_{N_1}$ , $5M_{N_1}$] \\   
		$M_{N_{2}}$ [GeV] & [$M_{N_1}$ , $10M_{N_1}$] \\
		$M_{N_{3}}$ [GeV] & [$M_{N_2}$ , $10 M_{N_2}$] \\
        $\lambda_3$  & [$10^{-3}$ , 1] \\
        $\lambda_4$  & [$-10^{-3}$,-1 ] \\
        $\lambda_5$  & [$10^{-11}$, $10^{-7}$] \\        
		\hline
	\end{tabular}
	\caption{Parameter ranges used in the numerical scan to identify the viable region of the model's parameter space.}
	\label{Table:Scan}
\end{table}
\section{Constraints}
\label{sec:constraints}
To ensure the theoretical consistency and phenomenological viability of our model, we impose a comprehensive set of theoretical and experimental constraints. We perform a detailed scan of the parameter space by varying the model parameters within the ranges listed in Table~\ref{Table:Scan}. We first delineate the allowed parameter space using theoretical and experimental constraints. Once the viable regions are identified, we evaluate their implications for dark matter production and explore the prospects for future detection.

\subsection{Theoretical Constraints}
\label{Sec:Theoritical_Constraints}

The model parameters must satisfy a set of theoretical requirements to ensure a consistent and stable scalar potential. We summarize these constraints below:

\begin{itemize}

	\item \textbf{\underline{Vacuum Stability:}} The scalar potential must be bounded from below to avoid instabilities. We apply the following necessary conditions derived in Ref.~\cite{Belyaev:2016lok}:
	\begin{align}
		\lambda_{1} > 0, \quad \lambda_{2} > 0, \quad \lambda_{3} + \sqrt{\lambda_1 \lambda_2} > 0, \quad \lambda_{3} + \lambda_{4} - |\lambda_5| + \sqrt{\lambda_1 \lambda_2} > 0.
	\end{align}

	\item \textbf{\underline{Perturbativity:}} To preserve the validity of perturbative quantum field theory, we require all quartic couplings to remain below $4\pi$. Additionally, one-loop corrections to the scalar couplings must not dominate over tree-level terms~\cite{Belanger:2022qxt, Garcia-Cely:2015khw}. The relevant conditions are:
	\begin{align}
		&\lambda_{2} < \frac{2\pi}{3}, \quad |\lambda_{3}| < 4\pi, \quad |\lambda_{4}| < 4\pi, \quad |\lambda_{5}| < 4\pi, \\
		&|\lambda_{3} + \lambda_{4}| < 4\pi, \quad |\lambda_{4} \pm \lambda_{5}| < 8\pi, \quad |\lambda_{3} + \lambda_{4} \pm \lambda_{5}| < 4\pi.
	\end{align}

	\item \textbf{\underline{Perturbative Unitarity:}} All $2 \to 2$ scalar scattering amplitudes must respect unitarity. The resulting constraints are given by~\cite{Arhrib:2012ia}:
	\begin{align}
		&|\lambda_{3} \pm \lambda_{4}| < 8\pi, \quad |\lambda_{3} \pm \lambda_{5}| < 8\pi, \quad |\lambda_{3} + 2\lambda_{4} \pm 3\lambda_{5}| < 8\pi, \\
		&|-\lambda_1 - \lambda_2 \pm \sqrt{(\lambda_1 - \lambda_2)^2 + 4\lambda_4^2}| < 16\pi, \\
		&|-3\lambda_1 - 3\lambda_2 \pm \sqrt{9(\lambda_1 - \lambda_2)^2 + 4(2\lambda_3 + \lambda_4)^2}| < 16\pi, \\
		&|-\lambda_1 - \lambda_2 \pm \sqrt{(\lambda_1 - \lambda_2)^2 + 4\lambda_5^2}| < 16\pi.
	\end{align}

	\item \textbf{\underline{Global Minimum Condition:}} We impose that the inert vacuum corresponds to the global minimum of the scalar potential~\cite{Ginzburg:2010wa}. Defining the parameter
	\begin{align}
		R = \frac{\lambda_L}{\sqrt{\lambda_1 \lambda_2}},
	\end{align}
	we require the following conditions to be satisfied:
	\begin{align}
		&m_\eta^2 > R \sqrt{\frac{\lambda_1}{\lambda_2}} m_\Phi^2 \quad \text{if } |R| < 1, \\
		&m_\eta^2 > \sqrt{\frac{\lambda_1}{\lambda_2}} m_\Phi^2 \quad \text{if } R > 1.
	\end{align}

	Moreover, following Ref.~\cite{Belyaev:2016lok}, we impose the additional condition:
	\begin{align}
		\lambda_4 - |\lambda_5| < 0,
	\end{align}
	to avoid charge-breaking vacua.
\end{itemize}

\subsection{Experimental Constraints}\label{sec:Experimental_Constraints}
In addition to the theoretical limits discussed previously, several experimental observations impose stringent constraints on the parameter space of the model. Note that, since the masses of the inert scalars (CP-even, CP-odd, and charged) are always greater than 100 GeV, constraints from LEP data do not apply in our scenario. Furthermore,  the experimental constraints described below will be applied sequentially and cumulatively to establish the model’s allowed parameter space.
\begin{itemize}

    \item \textbf{\underline{Electroweak precision data}:}
    Electroweak precision tests (EWPT) are characterized by three measurable parameters—S, T, and U—that encapsulate the effects of new physics beyond the SM on electroweak radiative corrections. In our analysis, we computed the S and T parameters using the approach described in Ref.~\cite{Barbieri:2006dq,Belyaev:2016lok}. According to the global electroweak fit~\cite{Lu:2022bgw}, the measured values for the $S$ and $T$ parameters are
    \begin{equation}
    S_{\rm exp} = 0.05 \pm 0.08, \qquad \qquad
    T_{\rm exp} = 0.09 \pm 0.07,
    \end{equation}
    assuming a SM Higgs mass of $m_h = 125~\text{GeV}$, $U = 0$, and a correlation coefficient $\rho_{ST} = +0.92$. The $\chi^2$ is defined as
    \begin{equation}
    \chi^2 =  \frac{1}{\left(1-\rho_{ST}^2\right)}\left[\frac{\left(S-S_{\rm exp}\right)^2}{\left( \Delta S_{\rm exp} \right)^2}+\frac{\left(T-T_{\rm exp}\right)^2}{\left( \Delta T_{\rm exp} \right)^2}-\frac{2 \rho_{ST} \left(S-S_{\rm exp}\right)\left(T-T_{\rm exp}\right)}{\Delta S_{\rm exp} \Delta T_{\rm exp}}\right],
    \end{equation}
    where $\Delta S_{\rm exp}$ and $\Delta T_{\rm exp}$ represent the $1\sigma$ experimental uncertainties. We select parameter points for which $\Delta \chi^2 = \chi^2 - \chi^2_{\text{min}} < 4$, corresponding to a confidence level of $95\%$.
	
	\item \textbf{\underline{Higgs Diphoton Rate ($h \to \gamma\gamma$)}:} The charged inert scalar contributes to the Higgs diphoton decay amplitude via loop diagrams. This contribution, proportional to $\frac{\lambda_3}{m_{\eta^\pm}^2}\mathcal{A}_0(\tau_{\eta^\pm})$ with $\tau_{\eta^\pm} = \frac{4m_{\eta^\pm}^2}{m_h^2}$, can either enhance or suppress the diphoton rate depending on whether it interferes constructively or destructively with the Standard Model (SM) contributions. The partial decay width of the SM-like Higgs boson to diphotons, including the charged scalar loop, is given by~\cite{Posch:2010hx, Krawczyk:2013jta}:
\begin{equation}
\Gamma(h \to \gamma\gamma) = \frac{G_F \alpha^2 M_h^3}{128\sqrt{2}\pi^3} \left| \sum_{f \in \{t,b,c,\tau\}} N_c Q_f^2 \mathcal{A}_{1/2}(\tau_f) + \mathcal{A}_1(\tau_W) + \frac{\lambda_3 v^2}{2M_{\eta^\pm}^2} \mathcal{A}_0(\tau_{\eta^\pm}) \right|^2,
\label{eq:hgaga}
\end{equation}
where $\tau_x = \frac{4m_x^2}{m_h^2}$, $Q_f$ are the fermion electric charges in units of the proton charge, and $N_c$ is the color factor. In the limit $\lambda_3 \to 0$, the expression reduces to the SM prediction: $\Gamma(h \to \gamma\gamma) = \Gamma^{\text{SM}}(h \to \gamma\gamma)$. The relevant loop functions can be found in Ref.~\cite{Swiezewska:2012eh}. The Higgs diphoton signal strength is defined as:
\begin{equation}
R_{\gamma\gamma} = \frac{\Gamma(h \to \gamma\gamma)}{\Gamma^{\text{SM}}(h \to \gamma\gamma)}.
\end{equation}

The most recent measurements of $R_{\gamma\gamma}$ from the ATLAS~\cite{ATLAS:2022tnm} and CMS~\cite{CMS:2021kom} collaborations are:
\begin{equation}
R_{\text{ATLAS}} = 1.04^{+0.10}_{-0.09}, \qquad 
R_{\text{CMS}} = 1.12 \pm 0.09.
\end{equation}

To combine these results, we perform a $\chi^2$ analysis following~\cite{Kraml:2019sis,Belanger:2024wca}. The combined $\chi^2$ is computed as:
\begin{equation}
\chi^2 = \sum_i \frac{(\mu_i - \hat{\mu})^2}{\sigma^+\sigma^- + (\sigma^+ - \sigma^-)(\mu_i - \hat{\mu})},
\label{chisq}
\end{equation}
where $\mu_i$ and $\sigma^\pm$ denote the experimental central values and uncertainties, and $\hat{\mu}$ is the predicted value from the model. We consider parameter points satisfying $\Delta\chi^2 = \chi^2 - \chi^2_{\text{min}} < 4$, corresponding to a 95\% confidence level.

\item \textbf{\underline{Charged Lepton Flavor Violation (cLFV)}:} The Yukawa interaction $Y^{\nu} \bar{L} \tilde{\eta} N$, responsible for neutrino mass generation, also induces lepton flavor-violating processes such as $\ell_\beta \to \ell_\alpha \gamma$ at one-loop level. The branching ratio for these processes is given by~\cite{Toma:2013zsa,Lindner:2016bgg}:
\begin{equation}
\text{Br}(\ell_\beta \to \ell_\alpha \gamma) = \frac{3 \alpha_{\rm em}}{64\pi G_F^2 m_{\eta^\pm}^4} \left| \sum_{i=1}^{3} Y^{\nu}_{\alpha i} Y^{\nu *}_{\beta i} F_2\left(\frac{M_{N_i}^2}{m_{\eta^\pm}^2}\right) \right|^2 \text{Br}(\ell_\beta \to \ell_\alpha \nu_\beta \bar{\nu}_\alpha),
\label{eq:BRmue2gamma}
\end{equation}
where $\alpha_{\rm em} = e^2 / 4\pi$ and $G_F$ is the Fermi constant. The loop function $F_2(x)$ can be found in Ref.~\cite{Toma:2013zsa}. The current experimental upper bounds on cLFV processes are~\cite{MEG:2016leq, ParticleDataGroup:2016lqr}:
\begin{align}
\text{Br}(\mu \to e\gamma) &\leq 4.3 \times 10^{-13}, \\
\text{Br}(\tau \to \mu\gamma) &\leq 4.4 \times 10^{-8},\\ 
\text{Br}(\tau \to e\gamma) &\leq 3.3 \times 10^{-8}.
\label{eq:lfv_expt_bound}
\end{align}
\end{itemize}

Similarly, branching ratio for the three-body decay processes $l_{\alpha} \rightarrow 3l_{\beta}$ is expressed as \cite{Liu:2022byu,Guo:2020qin},
\begin{align}
    \text{BR}(l_{\alpha}\rightarrow 3l_{\beta} ) 
    &= \frac{3 (4\pi)^2 \alpha^2}{8 G_F^2} 
       \Bigg[ |A_{ND}|^2 + |A_D|^2 
       \left( \frac{16}{3}\log\!\left(\frac{m_{\alpha}}{m_{\beta}}\right) - \frac{22}{3} \right) 
       + \frac{1}{6}|B|^2  \notag \\
    &\quad + \left( -2 A_{ND} A_{D}^* + \tfrac{1}{3} A_{ND} B^* - \tfrac{2}{3} A_{D} B^* + h.c. \right) \Bigg] 
    \text{BR}(l_{\alpha} \rightarrow l_{\beta} \nu_{\alpha} \bar{\nu}_{\beta}) \, .
\end{align}

Here, $A_{ND}$ and $A_D$ denote the non-dipole and dipole form factors, respectively, while $B$ represents the contribution from the box diagram. Their explicit forms can be found in Ref.~\cite{Liu:2022byu}. Among these three-body decay channels, the $\mu \rightarrow 3e$ process imposes the most stringent constraint on the parameter space considered in our analysis. The current experimental bound on the corresponding branching ratio is \cite{SINDRUM:1987nra} 
\[
    \text{BR}(\mu \rightarrow 3e) < 1.2 \times 10^{-12} \, .
\]
\begin{figure}[]
	\begin{center}
		\mbox{
		\subfigure[]{\includegraphics[height=6.5cm,width=8.00cm]{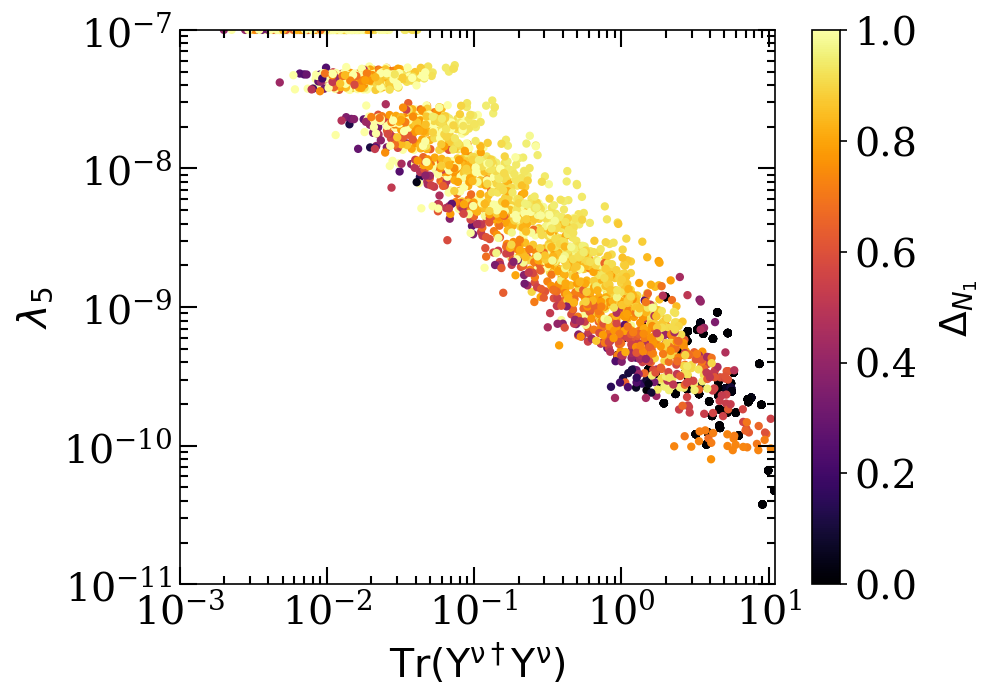}\label{fig:Relic_Allowed_Scan}} 
        \subfigure[]{ \includegraphics[height=6.5cm,width=8.00cm]{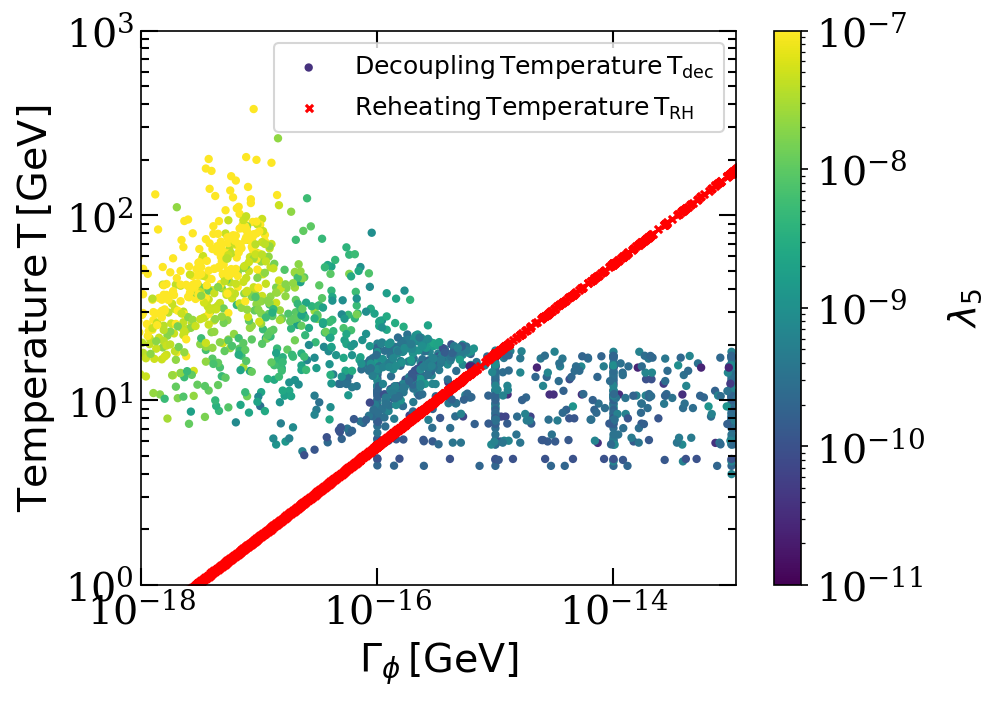}\label{fig:Temperature_Scan}}
		}
	\caption{Parameter space of the scotogenic model in the $\text{Tr}(Y^{\nu\dagger}Y^{\nu})$–$\lambda_5$ plane (left) and the $\Gamma_{\phi}$–$T$ plane (right), consistent with all theoretical and experimental constraints discussed in Section~\ref{sec:constraints}. The color scale corresponds to $\Delta_{N_1}$ in the left panel and to $\lambda_5$ in the right panel. Note that all points are consistent with the best-fit neutrino oscillation data and the observed DM relic density within the $2\sigma$ limit.}
		\label{fig:Gobal_scan_DM_Relic}
	\end{center}
\end{figure}
\section{Results and Discussion}
\label{sec:results}
\begin{figure}[]
	\begin{center}
       	\mbox{
		\subfigure[]{\includegraphics[height=6.5cm,width=8.0cm]{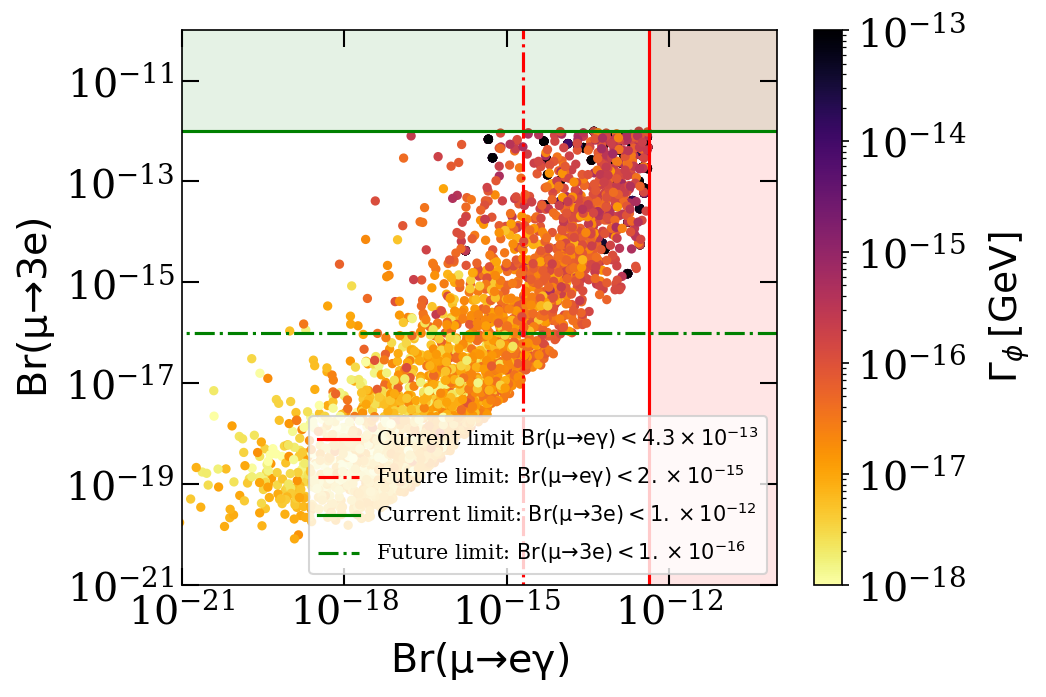}\label{fig:LFV_Scan}} 
        \subfigure[]{ \includegraphics[height=6.5cm,width=7.50cm]{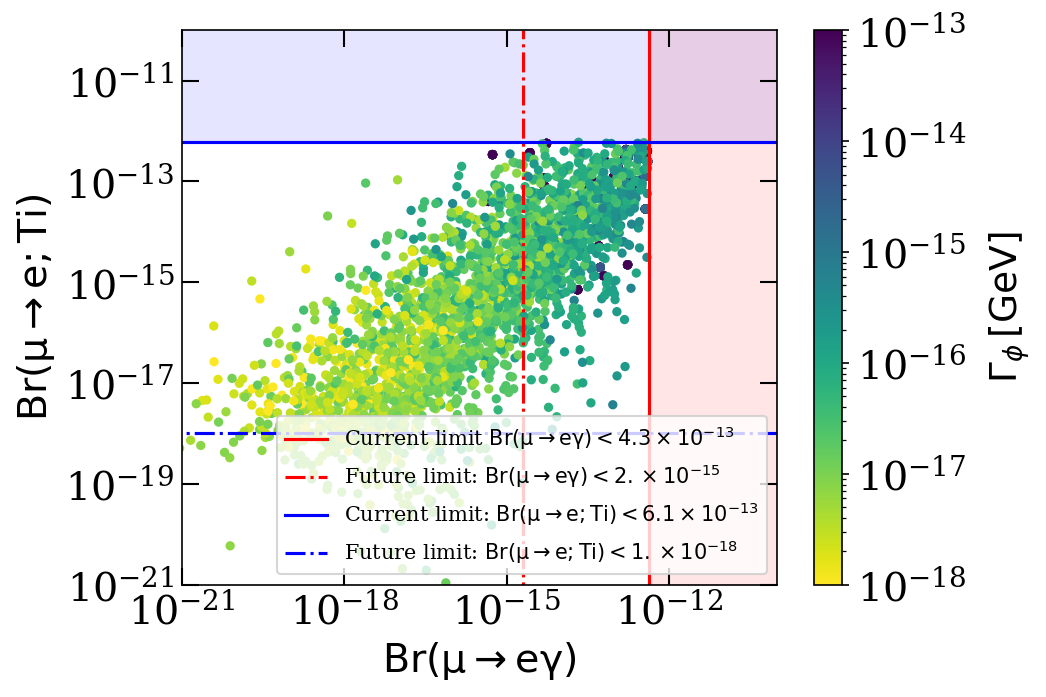}\label{fig:LFV_Scan2}}
		}  
	\caption{Left: Theoretically and experimentally allowed parameter points of the scotogenic model in the $\text{Br}(\mu \to e\gamma)$–$\text{Br}(\mu \to 3e)$ plane. The color scale denotes the inflaton decay width $\Gamma_{\phi}$. Also shown are the current limits from $\mu \to e\gamma$\cite{MEG:2016leq} (solid red) and $\mu \to 3e$\cite{SINDRUM:1987nra} (solid green), together with the projected sensitivities from $\mu \to e\gamma$\cite{MEGII:2018kmf} (dashed red) and $\mu \to 3e$\cite{Mu3e:2020gyw, Blondel:2013ia} (dashed green). Right: Scatter plot in the $\text{Br}(\mu \to e\gamma)$–$\text{Br}(\mu \to e;\rm{Ti})$ plane after imposing the current limits from $\text{Br}(\mu \to e;\rm{Ti})$ (solid blue)~\cite{Wintz:1998rp}, together with the projected sensitivity~\cite{prism}, shown as the dashed blue line.
    Note that all points are consistent with the best-fit neutrino oscillation data and the observed DM relic density within the $2\sigma$ limit.}
		\label{fig:Gobal_scan_LFV}
	\end{center}
\end{figure}
For all points satisfying the theoretical and experimental constraints obtained from the flat random scan in Table~\ref{Table:Scan}, we determined the value of $\Gamma_{\phi}$ consistent with the observed DM relic density within the $2\sigma$ limit. We again re-emphasize, as discussed in Section~\ref{sec:DMobs}, that when the DM undergoes early chemical decoupling during the reheating era, the DM number density is diluted by the entropy injection resulting from the inflaton decay. Thus, one can satisfy the observed DM relic density for $\langle\sigma v\rangle$ much lower than $\langle\sigma v\rangle\approx 3\times 10^{-26}\,\rm{cm^{3}sec^{-1}}$ which corresponds to the standard freeze-out mechanism. To delineate the parameter space corresponding to low and high reheating DM scenarios, we have defined the variable $\Delta_{N_1}$, which is given by,
\begin{equation}
    \Delta_{N_1}=1-\frac{\Omega^{obs}_{N_1}h^{2}}{\Omega_{N_1}h^{2}(T_{dec}<T_{RH})},
\end{equation}
where $\Omega_{N_1}h^{2}(T_{dec}<T_{RH})$ corresponds to the scenario in which $N_1$ undergoes chemical decoupling during the radiation-dominated era, followed by its freeze-out. In other words, it corresponds to instantaneous reheating scenario. Note that in the limit $\Delta_{N_1} \to 0$, the standard freeze-out result corresponding to instantaneous reheating is obtained. Conversely, in the scenario where $T_{dec}>>T_{RH}$, the limit $\Delta_{N_1} \to 1$ is recovered.

In Fig.~\ref{fig:Relic_Allowed_Scan}, we show the points in $\rm{Tr}(Y^{\nu\dagger}Y^{\nu})-\lambda_5$ plane with color pallet corresponding to $\Delta_{N_1}$. We clearly observe that as $\lambda_5$ decreases, $\mathrm{Tr}(Y^{\nu\dagger}Y^{\nu})$ increases. Consequently, the thermally averaged cross section $\langle\sigma v\rangle$ also increases with decreasing $\lambda_5$. The black points corresponding to $\Delta_{N_1} \to 0$ are obtained for $\lambda_5 \simeq \mathcal{O}(10^{-10} - 10^{-11})$, where the standard freeze-out result is recovered. In contrary, for $\lambda_5>4\times10^{-10}$, $\Delta_{N_1}$ deviates from $0$ and can reach to its maximum value of unity. This behavior is further illustrated in Fig.~\ref{fig:Temperature_Scan}, which shows the allowed points in the $\Gamma_{\phi}$–$T$ plane, with the color palette indicating the values of $\lambda_5$. We can see that for $\lambda_5>\mathcal{O}(10^{-10})$, $T_{dec}$ is much greater than $T_{\rm{RH}}$, which corresponds to $\Gamma_{\phi}<10^{-15}\,\rm{GeV}$. This arises from the fact that the DM abundance must be diluted by entropy injection in order to compensate for the decrease in $\langle\sigma v\rangle$. It is important to bring attention of the reader's that the DM primarily annihilates to second and third family of leptons, due to the hierarchical Yukawa structure, i,e $Y^{\nu}_{1e}<<Y^{\nu}_{1\mu}<Y^{\nu}_{1\tau}$, which arises from the cLFV constraints. Furthermore, the DM annihilation modes are insensitive to value of $\Gamma_{\phi}$. Hence, in our analysis, the DM $N_1$ behaves as $\mu\tau$-philic particle.

\begin{figure}[]
	\begin{center}
		\mbox{
		\subfigure[]{\includegraphics[height=6.5cm,width=8.0cm]{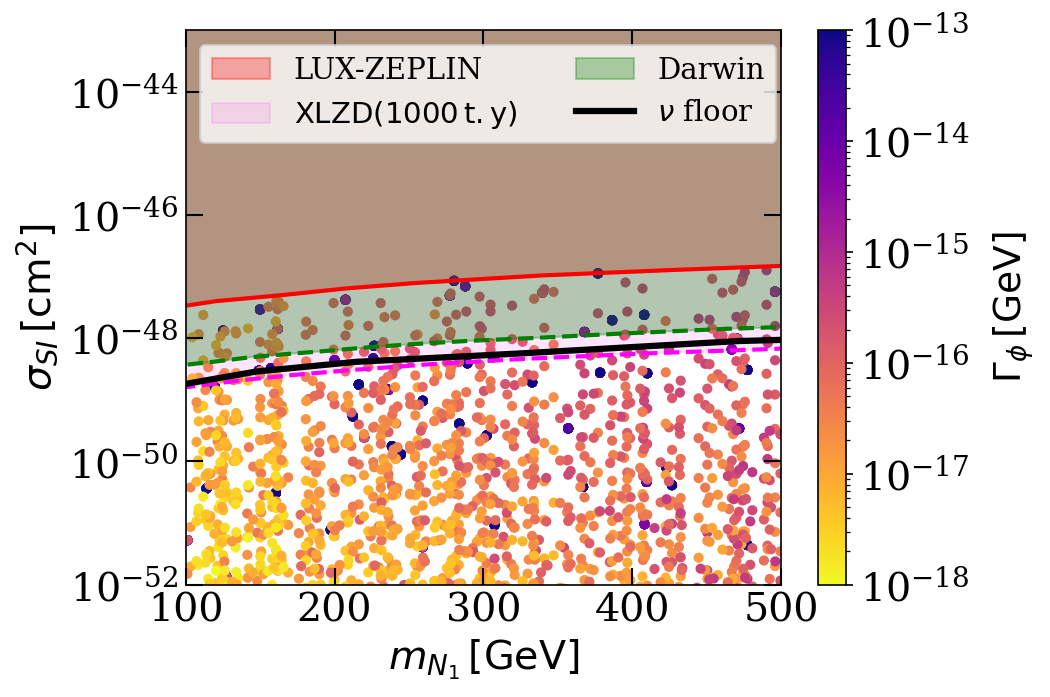}\label{fig:DD_Scan1}} 
        \subfigure[]{ \includegraphics[height=6.5cm,width=7.50cm]{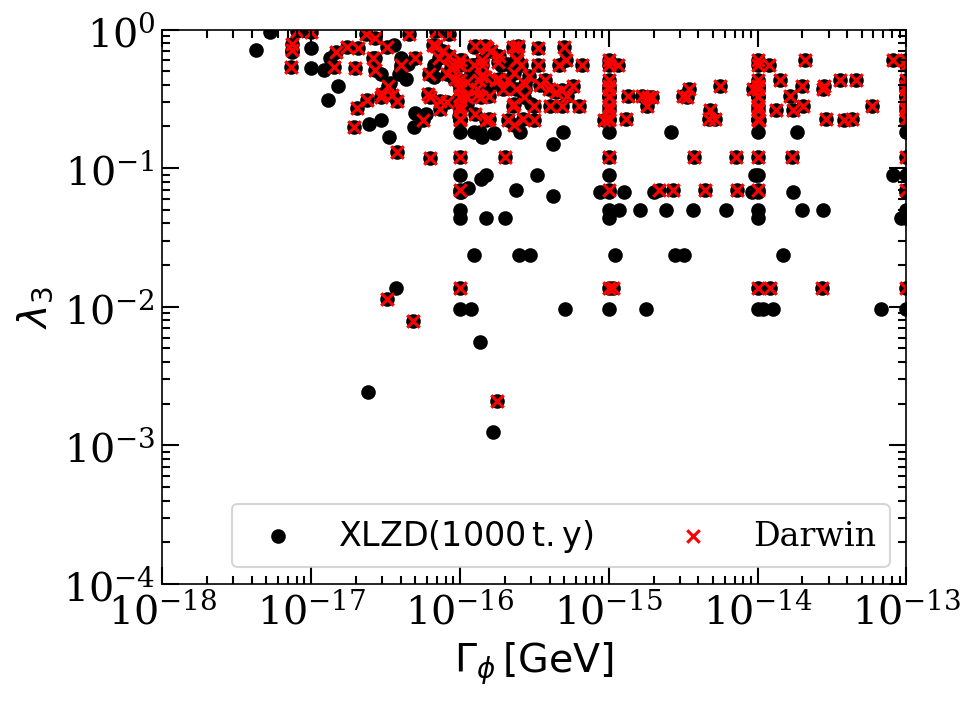}\label{fig:DD_Scan2}}
		}    
	\caption{Left: Theoretically and experimentally allowed parameter points of the scotogenic model in the $m_{N_1}$–$\sigma_{\text{SI}}$ plane. The color scale represents the inflaton decay width $\Gamma_{\phi}$. Right: Dependence of $\lambda_3$ on $\Gamma_{\phi}$, where red and black points correspond to regions that can be probed by the DARWIN~\cite{DARWIN:2016hyl} and XLZD~\cite{XLZD:2024nsu} direct detection experiments, respectively. Note that all points are consistent with the best-fit neutrino oscillation data, cLFV and the observed DM relic density within the $2\sigma$ limit.}
		\label{fig:Gobal_scan_DD}
	\end{center}
\end{figure}

In Fig.~\ref{fig:Gobal_scan_LFV}, we present the allowed parameter points in the $\mathrm{Br}(\mu\!\to\! e\gamma)$--$\mathrm{Br}(\mu\!\to\! 3e)$ plane (left panel) and the $\mathrm{Br}(\mu\!\to\! e\gamma)$--$\mathrm{Br}(\mu\!\to\! e;\mathrm{Ti})$ plane (right panel), where the colour scale denotes the value of $\Gamma_{\phi}$. Larger values of $\lambda_5$ suppress the neutrino Yukawa couplings and hence relax the cLFV constraints, whereas achieving the observed dark matter relic density in the low-reheating regime typically requires $\Gamma_{\phi}$ to decrease as $\lambda_5$ increases. This correlation is reflected in the scatter distributions: as $\Gamma_{\phi}$ decreases, the region with allowed points broadens in the cLFV branching-ratio planes. We further observe that the subset of points with $\Gamma_{\phi} \gtrsim 10^{-16}\,\mathrm{GeV}$ lies within the projected sensitivity of future $\mu\!\to\! e\gamma$~\cite{MEGII:2018kmf} and $\mu\!\to\! 3e$~\cite{Mu3e:2020gyw,Blondel:2013ia} searches, while parameter points with $\Gamma_{\phi} \gtrsim 10^{-17}\,\mathrm{GeV}$ fall within the expected reach of forthcoming $\mu\!\to\! e$ conversion experiments in titanium~\cite{prism}. Importantly, our analysis demonstrates that these next-generation cLFV facilities will be sensitive not only to the standard high-reheating scenarios but also to a substantial region of parameter space associated with low reheating temperatures, thereby providing an experimentally testable handle on this otherwise cosmology-driven region of the model.

In Fig~\ref{fig:DD_Scan1}, we show allowed points in $m_{N_1}-\sigma_{SI}$ plane, with color pallets corresponding to $\Gamma_\phi$.  The DM-nucleon spin-independent cross-section $\sigma_{SI}$ is primarily sensitive to the value of $\lambda_3$. It is observed that an increase in the value of $\lambda_3$ leads to an increase in $\sigma_{SI}$. In Fig.~\ref{fig:DD_Scan2}, we can see that for $\lambda_3>10^{-3}$,  the allowed points can be probed by the DARWIN~\cite{DARWIN:2016hyl} and 1000 tonne-year (t·y) liquid xenon exposures of the XLZD~\cite{XLZD:2024nsu} experiments. Note that in the left panel of Fig.~\ref{fig:DD_Scan2}, the allowed points span a wide range of $\Gamma_{\phi}$ values without showing a clear correlation. This behavior arises because $\Gamma_{\phi}$ primarily affects the dark matter relic density through its impact on the Yukawa coupling $Y^{\nu}$, whereas the spin-independent cross section $\sigma_{\text{SI}}$ depends mainly on $\lambda_3$ and is largely insensitive to $\Gamma_{\phi}$.

\section{Conclusion}
\label{sec:conclusion}

In this work, we have examined the phenomenology of fermionic dark matter in the scotogenic model, focusing on two complementary aspects: the role of direct detection experiments and cLFV searches, and the impact of a low reheating temperature on the relic density calculation. We have identified regions of the parameter space that satisfy the observed dark matter relic density and remain consistent with experimental searches, while excluding co-annihilations and fine-tuning of the Yukawa matrix responsible for neutrino mass generation. In these regions, the detection prospects are particularly promising in cLFV searches—especially $\mu \to 3e$ and $\mu\!\to\! e$ conversion, which offers greater sensitivity compared to $\mu \to e \gamma$. Upcoming searches for $\mu \to e\gamma$, $\mu \to 3e$, and $\mu\!\to\! e$ conversion are expected to test large portions of the viable parameter space. Furthermore, near-degenerate singlet fermions or sizable scalar quartic couplings can enhance the spin-independent scattering cross section to levels within the reach of future direct detection experiments such as DARWIN and XLZD.  

Beyond the standard thermal history, we have shown that scenarios with a low reheating temperature—where dark matter decouples before the end of reheating—can substantially alter the relic abundance through entropy injection. This effect reduces the annihilation cross section required to match the observed density, thereby enlarging the viable parameter space. Importantly, we demonstrated that the interplay between cLFV observables and direct detection remains robust even in this non-standard cosmological setting. Taken together, our findings illustrate that the fermionic dark matter scenario in the scotogenic model is far from inaccessible. Rather, it is poised to be tested comprehensively in the near future through the synergy of dark matter searches and cLFV experiments. 

\section*{Acknowledgements}
The work of AR is supported by the Basic Science Research Program through the National Research Foundation of Korea (NRF), funded by the Ministry of Education, through the Center for Quantum Spacetime (CQUeST) of Sogang University, with grant number RS-2020-NR049598, and by the Ministry of Science and ICT with grant number RS-2025-24523022.
\bibliographystyle{JHEP}
\bibliography{bibitem.bib}
\end{document}